# 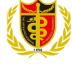 MambaClinix: Hierarchical Gated Convolution and Mamba-Based U-Net for Enhanced 3D Medical Image Segmentation

Chenyuan Bian[1*], Nan Xia[2], Xia Yang[1], Feifei Wang[1],

Fengjiao Wang[1], Bin Wei[1], and Qian Dong[1, 2, 3]

[1] Shandong Provincial Key Laboratory of Digital Medicine and Computer-Assisted Surgery,
The Affiliated Hospital of Qingdao University
`biancy@qdu.edu.cn`
[2] Institute of Digital Medicine and Computer Assisted Surgery, Qingdao University
[3] Shandong College Collaborative Innovation Center of Digital Medicine Clinical Treatment
and Nutrition Health, Qingdao University

**Abstract.** Deep learning, particularly convolutional neural networks (CNNs) and Transformers, has significantly advanced 3D medical image segmentation. While CNNs are highly effective at capturing local features, their limited receptive fields may hinder performance in complex clinical scenarios. In contrast, Transformers excel at modeling long-range dependencies but are computationally intensive, making them expensive to train and deploy. Recently, the Mamba architecture, based on the State Space Model (SSM), has been proposed to efficiently model long-range dependencies while maintaining linear computational complexity. However, its application in medical image segmentation reveals shortcomings, particularly in capturing critical local features essential for accurate delineation of clinical regions. In this study, we propose MambaClinix, a novel U-shaped architecture for medical image segmentation that integrates a hierarchical gated convolutional network (HGCN) with Mamba in an adaptive stage-wise framework. This design significantly enhances computational efficiency and high-order spatial interactions, enabling the model to effectively capture both proximal and distal relationships in medical images. Specifically, our HGCN is designed to mimic the attention mechanism of Transformers by a purely convolutional structure, facilitating high-order spatial interactions in feature maps while avoiding the computational complexity typically associated with Transformer-based methods. Additionally, we introduce a region-specific Tversky loss, which emphasizes specific pixel regions to improve auto-segmentation performance, thereby optimizing the model's decision-making process. Experimental results on five benchmark datasets demonstrate that the proposed MambaClinix achieves high segmentation accuracy while maintaining low model complexity. Our code is available at *https://github.com/CYB08/MambaClinix-PyTorch*.

## 1    Introduction

Medical image segmentation plays a pivotal role in enhancing diagnostic accuracy by precisely delineating anatomical structures [1, 2]. In this field, convolutional neural networks (CNNs) serve as the backbone due to their inherent translation invariance, which enables models to recognize identical patterns across different areas of an image [3-6]. This capability is crucial in medical imaging, where disease-related regions of interest may appear in various orientations and locations. However, the local receptive fields of CNNs restrict their ability to capture global features, posing challenges for medical image analysis [7, 8], where understanding larger anatomical structures and the spatial relationships between disparate parts of the clinical image is essential.

Transformer-based [9, 10] models have gained attention in medical image segmentation due to their ability to leverage the self-attention mechanism. This mechanism facilitates interactions between spatial features at various distances and complexities, surpassing the capabilities of standard CNNs and yielding impressive results [11-13]. However, despite their effectiveness in modeling extensive spatial relationships, Transformers require considerable computational resources, especially when dealing with large high-resolution 3D medical images, making them less suitable for lightweight model deployment [14]. Nonetheless, the self-attention mechanisms in Transformers offer valuable insights for advancing image segmentation models [12, 15]. Specifically, some studies [13, 16] indicate that the efficacy of the self-attention mechanism derives from its ability to capture high-order interactions within an image. This capability is achieved by dynamically adjusting attention weights and adaptively focusing across various spatial dimensions, which facilitates a comprehensive analysis of spatial feature relationships. Motivated by this, we have developed a novel Hierarchical Gated Convolutional Network (HGCN) for 3D medical image segmentation. Constructed with a purely convolutional structure, the HGCN recursively interacts with spatial features across multiple dimensions, effectively capturing and integrating multi-level spatial relationships, while avoiding the computational complexity associated with Transformer-based methods. However, further experiments have revealed diminishing cost-effectiveness in enhancing the spatial order of HGCN to capture long-range interactions as the network deepens. Additionally, computations involving higher-order spatial interactions require extensive recursive and gated convolution processes, which not only demand substantial GPU resources but also potentially increase the risk of overfitting.

Recently, State Space Models (SSMs) [17], particularly Mamba [18], have demonstrated a distinct advantage in efficiently capturing long-distance dependencies. Compared to Transformer-based models, Mamba's superior computational efficiency and hardware acceleration make it well-suited for processing large-scale data such as 3D medical images [19-21]. Following this, several studies [22, 23] have explored integrating CNNs with Mamba, aiming to combine CNNs' detailed local feature extraction with Mamba's global feature representation strength, thereby enhancing the performance of medical image segmentation.

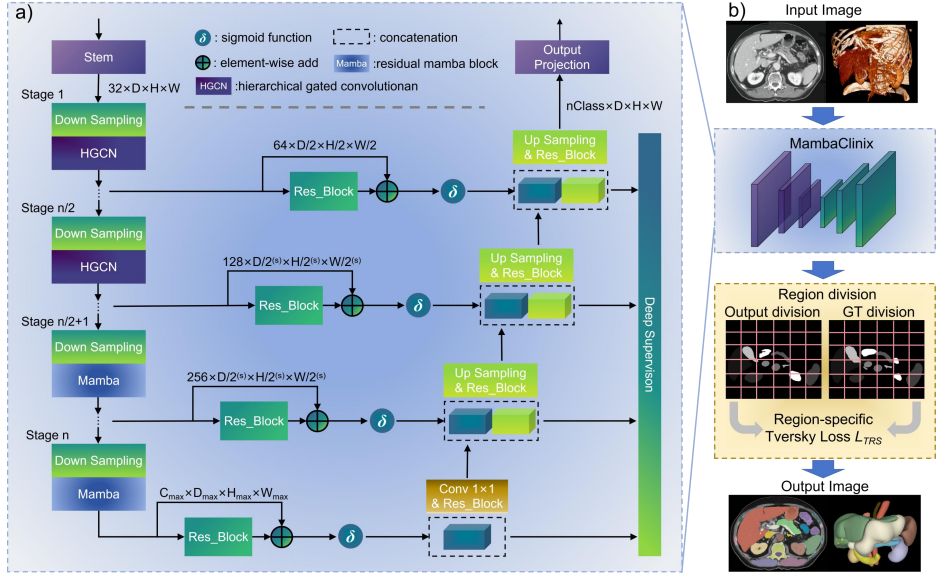

**Fig. 1.** Overview of the proposed MambaClinix. a) MambaClinix employs a stage-wise design with automatic configuration capabilities. It integrates Hierarchical Gated Convolutional Networks (HGCN) in early stages and Mamba blocks in later stages to enhance feature extraction across spatial scales. Deep supervision is conducted at each scale to ensure precision. b) MambaClinix applies a region-specific Tversky loss that divides the input image into sub-regions, with tailored loss calculations for each to refine segmentation outcomes.

In this paper, to balance our model while ensuring effective extraction of both local and global features, we propose MambaClinix, an adaptive stage-wise modeling framework that merges HGCN and Mamba, optimized for clinical image segmentation. In the lower stages of the encoder, HGCN is employed to extend high-order spatial interactions, capturing detailed elements that are crucial for precise medical segmentation. To mitigate the diminishing returns of increasing HGCN interaction orders, a residual Mamba module is incorporated in the higher stages to replace HGCN, extracting long-distance dependency features while maintaining computational efficiency. The HGCN enhances semantic features for the Mamba block, enriching the input quality. This stage-wise approach enables the model to achieve a deeper understanding of both proximal and distal relationships within medical images, essential for tasks such as organ segmentation and disease detection. Furthermore, MambaClinix inherits the self-configuring strategy from nnU-Net, allowing automatic adjustment of the network structure to match specific dataset characteristics. This adaptive configuration ensures that the architecture is finely tuned to meet the unique demands of diverse datasets, significantly boosting its clinical efficacy. Overall, our proposed MambaClinix integrates the high-order spatial interaction capabilities of CNNs with the global dependency extraction strengths of SSMs through an adaptive stage-wise design. This provides a flexible framework that seamlessly integrates into clinical medical image segmentation, minimizing manual intervention and enhancing the automation of diagnostic processes.

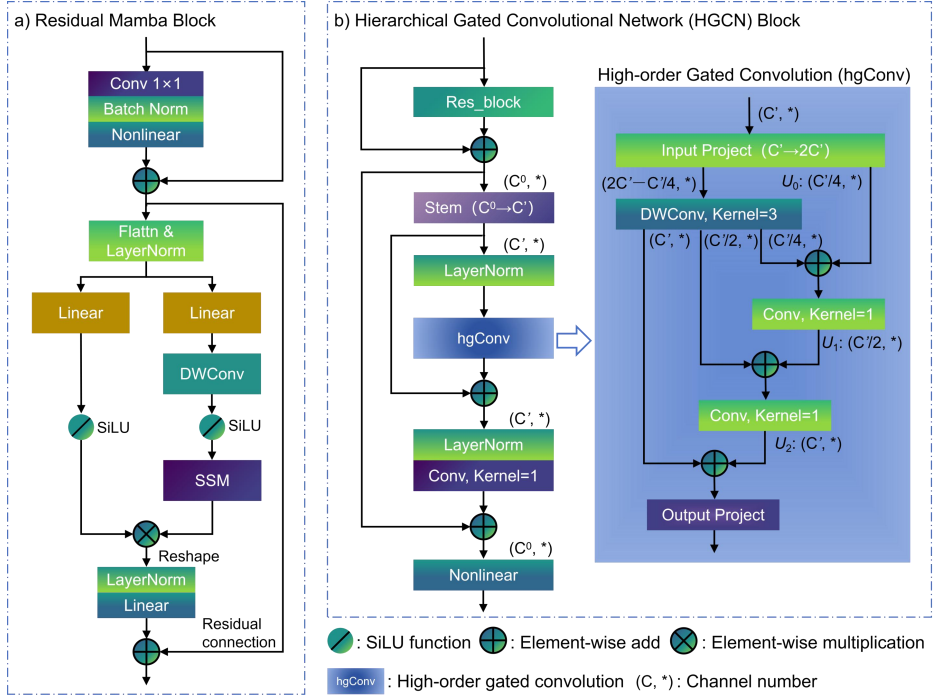

**Fig. 2.** a) The residual Mamba block, which incorporates residual connections and layer normalization. b) The Hierarchical Gated Convolutional Network (HGCN) block. This example illustrates an HGCN with 3-order spatial interactions, featuring a high-order gated convolution (hgConv) of order 3 implemented through three recursive steps.

## 2   Method

### 2.1   Architecture Overview

The proposed MambaClinix architecture, depicted in Fig. 1, employs an adaptive stage-wise design that combines specialized blocks within a U-shaped framework to enhance performance. In the lower stages of the encoder, we design a Hierarchical Gated Convolutional Network (HGCN) block, as shown in Fig. 2(b), to capture complex spatial relationships in volumetric medical images through a purely convolution structure. In the higher stages, a residual Mamba block, as shown in Fig. 2(a), replaces the HGCN to enhance the extraction of long-range dependencies. MambaClinix thus divides the encoder into two parts: the first leverages HGCN for efficient spatial feature interactions, while the second integrates these features with the residual Mamba block to ensure a comprehensive understanding of larger spatial contexts. This architecture enables the network to balance detailed feature processing with global feature integration, optimizing both the depth and breadth of medical image segmentation. MambaClinix adopts the self-configuring approach from nnU-Net, which automatically adapts the network structure to suit the specific characteristics of each dataset. Additionally, our HGCN is deeply integrated into this adaptive configuration, allowing for dynamic adjustment of spatial interaction orders based on dataset-specific features. This flexibility ensures that the HGCN can increase the order adaptively to balance structural complexity and computational efficiency. A region-specific Tversky loss function is also applied during model training, differentially weighting the loss for each sub-region of the 3D medical image [24, 25]. This approach directs the model's attention to more challenging segmentation areas, such as infected regions and small nodules, thereby improving detail learning and enhancing overall segmentation performance.

### 2.2 Hierarchical Gated Convolutional Network (HGCN)

The HGCN embedded within the framework captures image features dynamically in the early stages. Specifically, its core processing is handled by a high-order gated convolution (*hgConv*) layer, which employs gated convolutional mechanisms and recursive processes to facilitate efficient high-order spatial interactions. This HGCN bypasses the computational burden of self-attention mechanisms by extending interactions to multiple levels, enhancing the CNN's ability to capture multi-level spatial dependencies in medical images. Moreover, the *hgConv* adapts to various clinical data scenarios, seamlessly integrating with the adaptive framework. In the $s^{th}$ stage of the encoder, the input feature $h_{in}^{C^0}$ with $C^0$ channels first passes through a standard residual network (*Res*), followed by a stem layer (*Stem*) that adjusts the channel to $C'$. The *Stem* ensures a stable input structure for the subsequent *hgConv* operations while keeping them within the GPU's operational limits. After a layer normalization (*LN*), the $h_{in}^{C'}$ is processed through the *hgConv* to computes high-order spatial interactions. The resulting feature map $h_{out}^{C'}$ is then projected and combined with the residual component. The process can be represented as:

$$h_{in}^{C'} = Stem\left(Res(h_{in}^{C^0})\right) \tag{1}$$

$$h_{out}^{C'} = hgConv^{(s)}(LN(h_{in}^{C'})) + h_{in}^{C'} \tag{2}$$

$$h_{out}^{C^0} = \sigma(Conv\left(LN(h_{out}^{C'})\right) + Res(h_{in}^{C^0})) \tag{3}$$

**High-order Gated Convolution (*hgConv*).** Given a scaled 3D input feature $h^{C'}$ with $C'$ channels. The function $\varphi$ involves two sequential convolution processes to transform $h^{C'}$ into a channel-expanded and refined set of sub-vectors. The input projection first expends the number of channels from $C'$ to $2C'$, and then, following a depth-wise convolution function $f_{dw}$, the channels are further decomposed into various sub-vector combinations, which can be represented as:

$$\varphi_{dw}^{C' \to 2C'} : f_{dw}(h^{C' \to 2C' = C'_0 + \sum_{0 \leq j \leq n-1} C'_j}) = [U_0^{C'_0}, V_0^{C'_0}, ..., V_j^{C'_j}, ..., V_{n-1}^{C'_{n-1}}] \tag{4}$$

where, *n* represents the spatial interaction order of the *hgConv*. This subdivision is achieved by partitioning the channel $2C'$ into progressively smaller parts, as described in Eq. (5), enabling a hierarchical reduction in dimensionality. This approach promotes efficient data processing across the network's layers.

$$C'_j = \frac{C'}{2^{n-j-1}}, \quad 0 \leq j \leq n-1 \tag{5}$$

Each subdivided sub-vector then undergoes a transformation process from *j* to *j*+1, formulated as:

$$U_{j+1} = \begin{cases} V_j + \Phi_j(U_j) \times \gamma_f, & j = 1,...,n-1 \\ V_j + U_j, & j = 0 \end{cases} \tag{6}$$

where $\Phi_j$ represents a convolution function specific to *j*, and an externally provided parameter $\gamma_f$ modulates the previous output $U_j$. This operation enriches spatial information across different layers, enhancing the model's ability to capture complex patterns inherent in medical images.

### 2.3    Residual Mamba Block

The residual Mamba block utilizes residual connection and layer normalization to enhance the spatial modeling capabilities of the original Mamba. It is placed in the higher stages of the encoder to summarize the global features output by the HGCN block, capturing long-distance dependencies in medical images. In the $s^{th}$ stage, the scaled feature $m_{in} \in \Re^{C \times D' \times H' \times W'}$ first passes through a residual convolution with a kernel size of 1×1×1, followed by batch normalization (BN) and a non-linear activation function $\sigma$. The transformed output is then flattened, transposed, normalized, and processed through the Mamba blocks with two parallel branches. One branch starts with a linear function on the layer-normalized feature $\tilde{m}_{in}$ (denoted as *LN*), followed by a depth-wise convolution (*dwConv*) and the *SiLU* function. This feature enhancement is supported by the state space model (*SSM*) layer, expanding the block's capacity to capture longitudinal dependencies. For the another branch without *SSM* layer, the $\tilde{m}_{in}$ pass though a linear function followed by the *SiLU* function. Finally, the features from both branches are combined using the Hadamard product, reshaped to their original shape, and reintegrated into the network through a residual connection. The process can be written as:

$$\tilde{m}_{in} = \sigma(BN(Conv(m_{in}))) + m_{in} \tag{7}$$
$$\overline{m}_1 = SSM(SiLU(dwConv(Linear(LN(\tilde{m}_{in}))))) \tag{8}$$
$$\overline{m}_2 = SiLU(Linear(LN(\tilde{m}_{in}))) \tag{9}$$

$$m_{out} = MLP(LN(\overline{m}_1 \cdot \overline{m}_2)) + \tilde{m}_{in} \tag{10}$$

where, $MLP(\cdot)$ is a multi-layer perception structure that enriches the representation of the merged features. This architecture preserves the continuity and integrity of the image feature flow, facilitating gradient propagation across the network.

### 2.4 Stage-wise Integration of HGCN and Mamba

As the network deepens, the benefits of increasing HGCN's computational orders to capture global features in medical images diminish. To address this, we introduce the residual Mamba block in the higher stages, replacing HGCN. This substitution is designed to extract long-range dependencies while preserving computationally efficiency. The detailed image features captured by HGCN in the earlier stages provide enriched information to the Mamba block, enhancing the quality of the input it receives. Given an encoder with $s$ stages, where $\mathcal{H}$ and $\mathcal{M}$ represents the HGCN and residual Mamba block, respectively. The stage-wise encoder $\Sigma_e$ can be formulated as:

$$\Sigma_e = [\mathcal{H}^1_{(2)}, \mathcal{H}^2_{(3)}, \ldots, \mathcal{H}^{s//2}_{(s//2+1)}, \mathcal{M}^{s//2+1}, \ldots, \mathcal{M}^{s-1}, \mathcal{M}^s] \tag{11}$$

where, $s$ is the total number of stages and is adaptively adjusted based on the specific dataset characteristics. To account for computational complexity and GPU limits, the HGCN block is configured to compute a minimum spatial interaction order of 2 and a maximum of 6 by default. $\mathcal{H}^{s//2}_{(s//2+1)}$ denotes the HGCN block in the $s//2^{th}$ stage, which computes spatial interactions of order ($s//2+1$); $\mathcal{M}^s$ represents the Mamba block in the $s^{th}$ stage. This illustrates how the customizable capability of $\mathcal{H}^{s//2}_{(s//2+2)}$ is deeply integrated with our adaptive configuration strategy, allowing it to adjust its structure across different stages to accommodate varying data scenarios. An example with 6 stages, comprising three HGCN blocks and three Mamba blocks, is given as follow:

$$\tilde{\theta}_0 = \mathcal{H}^1_{(2)}(ds_1(\theta_0)), \tilde{\theta}_1 = \mathcal{H}^2_{(3)}(ds_2(\tilde{\theta}_0)), \tilde{\theta}_2 = \mathcal{H}^3_{(4)}(ds_3(\tilde{\theta}_1)) \tag{12}$$

$$\tilde{\beta}_3 = \mathcal{M}^4(ds_4(\tilde{\theta}_2)), \tilde{\beta}_4 = \mathcal{M}^5(ds_5(\tilde{\beta}_3)), \tilde{\beta}_5 = \mathcal{M}^6(ds_6(\tilde{\beta}_2)) \tag{13}$$

where, $ds(\cdot)$ represents the down-sampling operation. The initial input $\theta_0$ undergoes HGCN processing by $\mathcal{H}^0_{(2)}$ after a down-sampling, resulting in $\tilde{\theta}_0$. This pattern of down-sampling followed by HGCN continues, with $\tilde{\theta}_0$ and subsequent layers processed through $\mathcal{H}^1_{(3)}$ and $\mathcal{H}^2_{(4)}$, respectively, leading to progressively refined outputs. The feature map is then passed through the Mamba block $\mathcal{M}^3$. The process continues with $\mathcal{M}^4$ and $\mathcal{M}^5$ after additional down-sampling steps, producing the final output $\tilde{\beta}_5$. This mode efficiently utilizes the network's structure, optimizing computational processes and enhancing its ability to capture complex spatial interactions, which is crucial for tasks such as medical image segmentation.

### 2.5   Region-Specific Loss Function

In medical image analysis, the Dice loss function is a well-established metric [26, 27] for evaluating segmentation model performance. This function is defined as:

$$L_{Dice} = 1 - \frac{2\sum_i^N \hat{y}_i y_i + \varepsilon}{\sum_i^N \hat{y}_i + \sum_i^N y_i + \varepsilon} \tag{14}$$

where $\hat{y}_i$ represents the predicted probability for voxel $i$, an $y_i$ corresponds to the ground truth binary values for voxel $i$. The sums are taken over all $N$ voxels in the image. The smoothing term $\varepsilon$ prevents division by zero and ensures computational stability. During training, the Dice loss function optimizes true positives (TPs) while equally penalizing false positives (FPs) and false negatives (FNs). However, the disparity in volumetric medical images between target organs and the background often results in data imbalance, biasing predictions toward the background and increasing FNs more than FPs. In clinical settings, high recall (reducing FNs) is prioritized to ensure every potential issue is detected, even at the cost of some FPs. For instance, missing a small nodule in a lung CT scan could delay lung cancer treatment, while an FP can be corrected with additional manual review. To address this, the Tversky loss function was developed [25, 28], which is defined as:

$$L_T = 1 - \frac{\sum_i^N \hat{y}_{0i} y_{0i} + \varepsilon}{\sum_i^N \hat{y}_{0i} y_{0i} + \alpha \sum_i^N \hat{y}_{0i} y_{1i} + \beta \sum_i^N \hat{y}_{1i} y_{0i} + \varepsilon} \tag{15}$$

where $\hat{y}_{0i}$ is the probability of voxel $i$ belonging to a target class, and $\hat{y}_{1i}$ is the probability of voxel $i$ being a background. The hyperparameters $\alpha$ and $\beta$ control the trade-off between FPs and FNs, respectively. When $\beta$ is greater than $\alpha$, the Tversky loss places more emphasis on FN errors to improve recall. However, the Tversky loss function focus on total volume overlap, treating all voxels uniformly in both target and background regions [24]. This approach overlooks the diverse segmentation difficulties across different sub-regions. To overcome this limitation, we implemented a region-specific loss function that focuses on optimizing specific sub-regions within volumetric medical images. During training, this method dynamically adjusts penalties for different regions to boost overall prediction accuracy, assigning higher weights to sub-regions that are more challenging to segment. The region-specific Tversky loss function can be expressed as:

$$L_{TRS} = \sum_k (1 - \frac{\sum_i^{N_k} \hat{y}_{ic} y_{ic} + \varepsilon}{\sum_i^{N_k} \hat{y}_{ic} y_{ic} + \alpha \sum_i^{N_k} \hat{y}_{ic} y_{i\bar{c}} + \beta \sum_i^{N_k} \hat{y}_{i\bar{c}} y_{ic} + \varepsilon}) \tag{16}$$

where, a medical image volume consisting of $N$ voxels is partitioned into $k$ sub-volumes, denoted as $N = \{N_1, N_2, ..., N_k\}$. The model calculates distinct losses for these sub-volumes and the gradient of the region-specific loss depends solely on the evaluation of these sub-regions rather than the entire image volume.

## 3 Experiment and results

### 3.1 Datasets

To acquire data that reflects diverse and real-world clinical scenarios, we conducted a retrospective study at the Shandong Key Laboratory of Digital Medicine and Computer Assisted Surgery, The Affiliated Hospital of Qingdao University. This study collected three distinct datasets (**PCD, LungT,** and **LiverT**) from the years 2021 to 2023. Additionally, we evaluated the performance and scalability of MambaClinix using two public datasets (**ABD** [29] and **BraTs** [30]) encompassing various disease areas. Dataset details are provided in Table 1. Furthermore, MambaClinix inherits nnU-Net's self-configuration capability, enabling dynamic adjustments to suit specific dataset features. The network configuration parameters for these datasets are outlined in Table 2.

(1) **Pulmonary Circulatory CECT Dataset (PCD).** This in-house dataset comprises 547 3D contrast-enhanced CT (CECT) images, precisely annotated to segment the pulmonary circulation system, with labels for the *Bronchus*, *Pulmonary Artery*, and *Pulmonary Vein*. It was derived from randomly sampled cases at the Affiliated Hospital of Qingdao University between 2022 and 2023. Each image volume was carefully crafted by three experienced radiologists and later reviewed by clinical doctors to ensure accuracy. This dataset has been used for assisted diagnosis in clinical settings.

(2) **Lung Tumor CECT Dataset (LungT).** This in-house dataset includes 800 CECT image volumes focusing on lung tumors. It was collected from randomly sampled cases at the Affiliated Hospital of Qingdao University during 2021-2023. Annotations were carefully crafted by three experienced radiologists and later refined by clinical physicians to ensure utility in clinical diagnosis and surgical planning.

(3) **Liver Tumor CT Dataset (LiverT).** This dataset consists of 292 liver tumor samples, divided into two parts. The first part is an in-house collection of 161 CECT images, derived from randomly selected cases at the Affiliated Hospital of Qingdao University between 2021 and 2023. These sample were annotated by experienced physicians and validated by clinical doctors. The second part comprises 131 samples from the public Medical Segmentation Decathlon (MSD) challenge [31], specifically focused at liver tumor segmentation from CT images [32], with each image volume clearly labeled to distinguish between liver and tumors.

(4) **Abdomen CT Dataset (ABD)**. This dataset, derived from the MICCAI 2022 FLARE challenge, is a 3D multi-organ segmentation dataset comprised of 100 CT volumes. It is designed for segmenting 13 different abdominal organs. In alignment with the U-Mamba settings [23], the training set incorporates 50 CT volumes from the MSD Pancreas dataset, with annotations provided by AbdomenCT-1K [33]. An additional 50 cases, collected from various medical centers, serve as the testing set, with annotations furnished by the challenge organizers. This setup provides a robust assessment of models across diverse imaging scenarios.

(5) **Brain Tumor MRI Dataset (BraTS).** This dataset, sourced from the BraTS2021 challenge [34], is specifically designed for brain tumor segmentation and contains 1,251 3D brain Magnetic Resonance Imaging (MRI) volumes[30]. Each volume features four imaging modalities (T1, T1ce, T2, and Flair). The dataset targets three specific segmentation areas: Whole Tumor (WT), Enhancing Tumor (ET), and Tumor Core (TC).

**Table 1.** Summary of datasets. "Resampling" indicates the spatial resolution after preprocessing. "Tr/Va/Ts" denotes the data split into training (TR), validation (VA), and test sets (TS). "Labels" represents the number of target regions to be segmented.

| Dataset | Resampling | Tr/Va/Ts | Labels | Data Availability |
|---|---|---|---|---|
| PCD | (1.25, 0.77, 0.77) | 397/100/50 | 3 | In-house |
| LungT | (1.25, 0.77, 0.77) | 560/160/80 | 1 | In-house |
| LiverT | (1.22, 0.76, 0.76) | 209/53/30 | 1 | In-house and Public |
| ABD | (2.5, 0.80, 0.80) | 40/10/50 | 13 | Public |
| BraTS | (1.0, 1.0, 1.0) | 876/250/125 | 3 | Public |

**Table 2.** Self-configuration of MambaClinix according to specific dataset. For example, $[\mathcal{H},\mathcal{H},\mathcal{H},\mathcal{M},\mathcal{M},\mathcal{M}]$ denotes that the encoder's first three stages are equipped with HGCN blocks, and the last three stages utilize Mamba blocks.

| Datasest | Adaptive Configurations | | | | |
|---|---|---|---|---|---|
| | Stages | Encoder Blocks | Patch size | Batch size | Pooling per axis |
| PCD | 6 | $[\mathcal{H},\mathcal{H},\mathcal{H},\mathcal{M},\mathcal{M},\mathcal{M}]$ | (80, 192, 160) | 2 | (4, 5, 5) |
| LungT | 6 | $[\mathcal{H},\mathcal{H},\mathcal{H},\mathcal{M},\mathcal{M},\mathcal{M}]$ | (96, 160, 160) | 2 | — |
| LiverT | 6 | $[\mathcal{H},\mathcal{H},\mathcal{H},\mathcal{M},\mathcal{M},\mathcal{M}]$ | (64, 192, 192) | 2 | (4, 5, 5) |
| ABD | 6 | $[\mathcal{H},\mathcal{H},\mathcal{H},\mathcal{M},\mathcal{M},\mathcal{M}]$ | (40, 224, 192) | 2 | (3, 5, 5) |
| BraTS | 6 | $[\mathcal{H},\mathcal{H},\mathcal{H},\mathcal{M},\mathcal{M},\mathcal{M}]$ | (128, 128, 128) | 2 | — |

### 3.2 Implementation Setup

MambaClinix was implemented on PyTorch 2.0.1 based on the nnU-Net and UMamba frameworks. All experiments were conducted on four NVIDIA GeForce RTX 4090 GPUs. The Adam was employed as the optimizer, initialized with a learning rate of 1$e$-4. The *PolyLRScheduler* was used for learning rate scheduling. MambaClinix was trained for 300 epochs on the PCD dataset, 1000 epochs on the LungT dataset and the LiverT dataset, 500 epochs on the ABD dataset and the BraTS dataset.

**Table 3.** Benchmark results of comparative medical image segmentation methods, measured in terms of DSC and mIoU scores. These results are based on comparisons across five datasets, with bold values indicating the best DSC score. "Pam" denotes the number of trainable parameters.

| Methods | Pam | PCD | | LungT | | LiverT | | ABD | | BraTS | |
|---|---|---|---|---|---|---|---|---|---|---|---|
| | | DSC | mIoU | DSC | mIoU | DSC | mIoU | DSC | mIoU | DSC | mIoU |
| *CNN-based* | | | | | | | | | | | |
| nnU-Net | 89M | **86.48** | 77.91 | 71.51 | 60.78 | 66.57 | 55.99 | 84.03 | 76.18 | 90.38 | 82.75 |
| SegResNet | 19M | 85.43 | 76.76 | 62.37 | 50.89 | 68.58 | 57.74 | 78.50 | 69.60 | 90.68 | 83.03 |
| *Transformer-based* | | | | | | | | | | | |
| UNETR | 93M | 83.30 | 74.29 | 69.43 | 58.54 | 54.95 | 43.74 | 66.28 | 58.41 | 89.31 | 81.95 |
| SwinUNETR | 61M | 83.82 | 75.87 | 67.64 | 56.48 | 63.54 | 53.25 | 80.56 | 72.27 | 90.32 | 82.58 |
| *Mamba-based* | | | | | | | | | | | |
| U-Mamba Bot | 43M | 86.22 | 77.06 | 70.18 | 59.28 | 70.67 | 61.56 | 84.11 | 76.94 | **90.91** | 84.23 |
| LightM-UNet | 5M | 85.13 | 76.06 | 67.42 | 55.73 | 68.48 | 56.72 | 73.87 | 66.62 | 89.21 | 80.49 |
| MambaClinix | 49M | 86.37 | 77.53 | **72.78** | 61.67 | **71.37** | 61.75 | **84.69** | 77.15 | 90.77 | 83.85 |

### 3.3 Benchmark Results

To evaluate the efficacy of MambaClinix, we conducted a comparative analysis against several established models in the field of medical image segmentation. The results are presented in Table 3 and Fig. 3-5. All comparative models were categorized into three groups based on their architectural foundations: CNN-based models (e.g., nnU-Net [7], SegResNet [35]), Transformer-based models (e.g., UNETR [36], SwinUNETR [12]), and Mamba-based models (e.g., U-Mamba [23], LightM-Unet [19], and MambaClinix). Performance was assessed using metrics such as the Dice Similarity Coefficient (DSC) and the mean Intersection over Union (mIoU). All models were implemented using the nnU-Net framework, with consistent image preprocessing to ensure a fair comparison.

On the LungT dataset, MambaClinix achieved a Dice score of 72.78%, a notable result given the high variability and similar density of lung tissues, demonstrating its ability to accurately segment small, less-contrasted regions. On the ABD dataset, which contains a range of abdominal organs with complex geometries, MambaClinix achieved the highest Dice score of 84.69%. Notably, the training and testing sets for this dataset came from different data sources, highlighting the model's robustness in global feature extraction and key feature transfer. For the LiverT dataset, the data was derived from a blend of two distinct sources that were shuffled to ensure both the training and testing sets consisted of multi-source data. MambaClinix achieved a Dice score of 71.37%, outperforming other models and demonstrating its effectiveness in handling images with significant structural abnormalities and its adaptability to multi-source medical images. While MambaClinix did not achieve the highest scores on the PCD and BraTS datasets, it still produced highly competitive results, showcasing its robustness in complex clinical scenarios.

Furthermore, experimental results show that Mamba-based methods require fewer parameters compared to Transformer-based models, thereby improving computational efficiency. Although SegResNet and LightM-UNet utilize even fewer parameters, they exhibited significant performance declines when transferred to different data sources. For example, both models underperformed on the ABD dataset, where the test data came from different sources, likely due to their reduced parameter counts affecting model generalization. In contrast, MambaClinix, with an optimal balance of parameters, not only improves accuracy but also ensures robust generalization across diverse datasets.

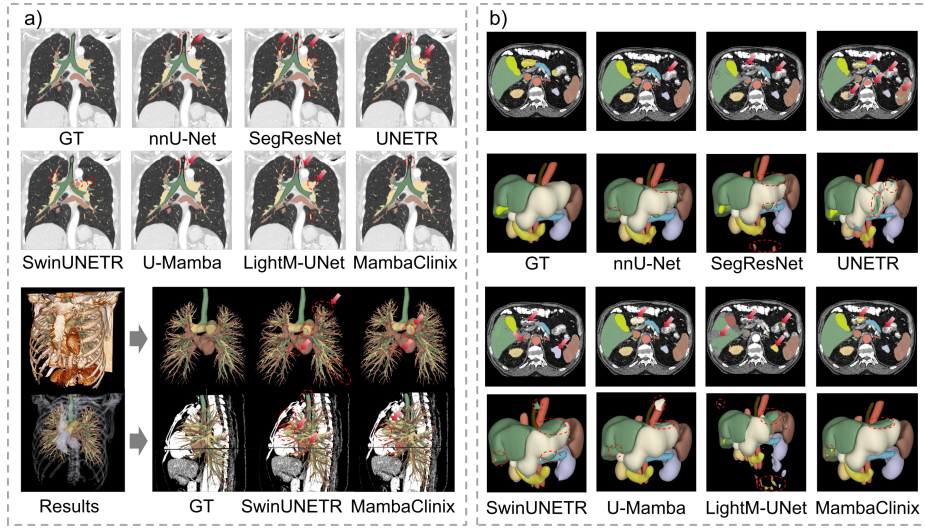

**Fig. 3**. Visual comparisons on a) PCD dataset and b) ABD dataset, highlighting areas with significant differences. The results show that MambaClinix can effectively recognize target positions and shapes.

### 3.4   Ablation Study

We conducted an ablation study to evaluate the efficacy of different blocks within the MambaClinix architecture. The experiments were performed on the BraTS, LiverT, and LungT datasets, with the results detailed in Table 4. The baseline model was established by replacing the specialized Mamba and HGCN blocks in MambaClinix with standard residual blocks. This setup served as a control for subsequent experiments, where various blocks were incrementally integrated or replaced into the baseline configuration. This approach allowed for a systematic evaluation of each block's contribution to overall segmentation performance.

Table 4. Ablation study for different modeling blocks, measured using DSC score [%]. FLOPs denotes floating point operations, with units in gigabytes [GB].

| Methods | Pam (M) | FLOPs (G) | BraTS | | | LiverT | LungT |
| --- | --- | --- | --- | --- | --- | --- | --- |
| | | | WT | TC | ET | TC | TC |
| Baseline | 14.37 | 995.217 | 88.43 | 87.57 | 84.36 | 48.64 | 61.07 |
| + only Mamba | 16.61 | 1057.89 | 90.40 | 89.63 | 87.19 | 67.47 | 63.12 |
| + only HGCN[2, 3, 4, 5, 6, 6] | 24.21 | 2128.49 | 92.46 | 91.51 | 87.68 | 70.47 | 69.49 |
| + only HGCN[3, 3, 3, 3, 3, 3] | 24.21 | 2130.02 | 91.99 | 91.46 | 87.21 | 68.24 | 69.85 |
| + only HGCN[4, 4, 4, 4, 4, 4] | 24.22 | 2130.74 | 92.15 | 91.65 | 88.36 | 69.87 | 70.40 |
| + HGCN[2, 3, 4, 5, 6] + Mamba (Bot) | 21.67 | 2127.67 | 92.27 | 91.48 | 87.74 | 67.25 | 70.48 |
| + HGCN[4, 4, 4] + Mamba (Stage-wise) | 17.41 | 2087.32 | 92.54 | 91.43 | 86.28 | 70.58 | 71.67 |
| + HGCN[4, 5, 6] + Mamba (Stage-wise) | 17.42 | 2088.36 | **92.73** | 91.70 | 88.03 | 71.13 | 71.83 |
| + HGCN[2, 3, 4] + Mamba (Stage-wise) | 17.41 | 2085.79 | 92.02 | **91.70** | **88.60** | **71.37** | **72.78** |

(1) **Mamba Effectiveness**: Incorporating the Mamba block into the baseline model significantly improved performance, particularly in the Enhancing Tumor (ET) segmentation task on the BraTS dataset, where the DSC score increased by 2.83. This improvement is likely due to Mamba's ability to capture long-range features, enhancing target identification accuracy.

(2) **HGCN Effectiveness:** To assess the impact of using HGCN alone (without Mamba), we constructed three distinct encoder configurations with varying interaction orders. These configurations were: 1) HGCN[2,3,4,5,6,6], featuring progressively increasing HGCN orders; 2) HGCN[3,3,3,3,3,3], with uniform HGCN blocks of order 3; and 3) HGCN[4,4,4,4,4,4]: with all stages set to order 4. These setups were assessed to identify the optimal structure. The HGCN blocks with progressively increasing orders showed superior computational efficiency, with lower FLOPs. In particular, for the WT segmentation task in the BraTS dataset, HGCN[2,3,4,5,6,6] achieved a Dice score of 92.46, outperforming the HGCN[3,3,3,3,3,3] and HGCN[4,4,4,4,4,4] configurations, which scored 91.99 and 92.15, respectively. This progressively increasing order design proved advantageous, as simpler network structures in the early stages facilitated efficient feature extraction, while more complex interactions in later stages captured a broader range of global features, improving overall performance.

(3) **Effectiveness of Combining Mamba and HGCN**: We evaluated the impact of combining Mamba and HGCN blocks within the MambaClinix network. MambaClinix's adaptive module capability allows it to automatically adjust the number of stages in its U-shaped structure based on the dataset characteristics, enabling dynamic customization of the integration between HGCN and Mamba blocks. In the lower stages (the first three stages in this study), we used HGCN blocks with incrementally increasing orders, while the higher stages (the last three stages) employed Mamba blocks. For a thorough ablation study, we tested three different combinations of lower-stage HGCN blocks with varying spatial interaction orders: HGCN[4,4,4], HGCN[4,5,6], and HGCN[2,3,4]. Additionally, we evaluated a configuration where only the Mamba block was placed in the final stage as a bottleneck: HGCN[2,3,4] + Mamba (bot). The experimental results revealed that combining Mamba and HGCN blocks yields superior performance compared to using them individually. Except for the WT segmentation task in the BraTS dataset, the combined model consistently delivered the best results across all other tasks. Furthermore, this hybrid approach also demonstrates advantages in terms of parameter efficiency.

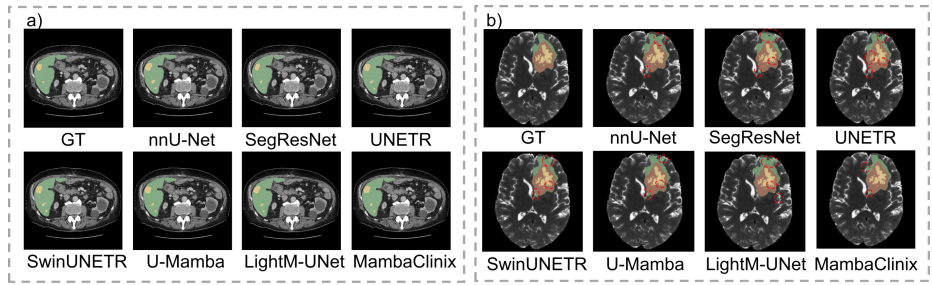

**Fig. 4.** Visual comparisons on a) LiverT dataset and b) BraTS dataset, highlighting areas with significant differences.

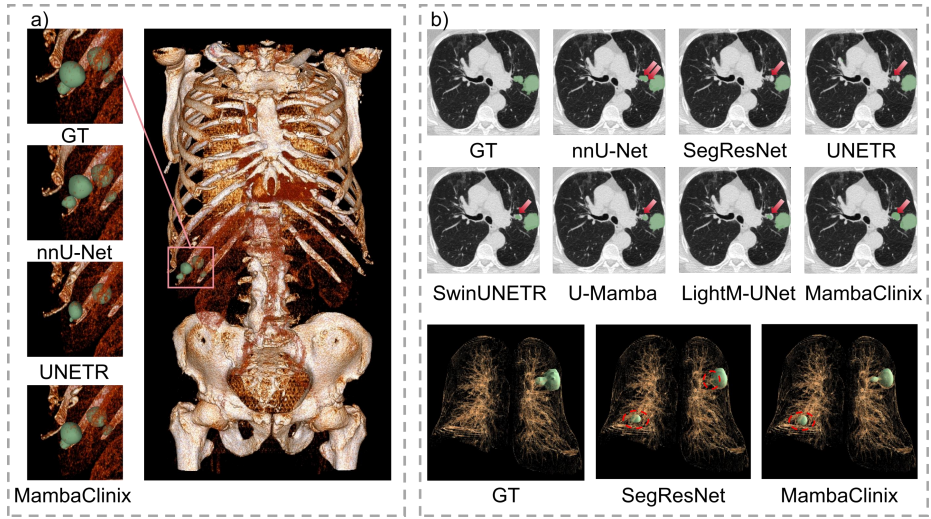

**Fig. 5.** Visualization of results. a) 3D visualization and detailed shapes of the liver tumors on the LiverT dataset;  b) Segmentation results of the LungT dataset, including the spatial positioning of lung tumors.

### 3.5   Efficiency of Region Specific Loss

In this experiment, the baseline was established using a compound loss function combining cross entropy (CE) and Dice. We further divided the image volume into sub-regions and applied region-specific Dice loss calculations to optimize predictions for each sub-region individually, refer to as *Baseline(RS)*. Next, a Tversky loss term ($L_{TRS}$) was incorporated, with fixed penalty coefficients for FP and FN set at 0.3 and 0.7, respectively. Finally, an adaptive strategy for these penalty coefficients was introduced, forming the final loss function for MambaClinix. The results are provided in Table 5.

**Table 5.** Ablation study for loss function. "Baseline" represents the compound loss of cross entropy (CE) and Dice. "Baseline (RS)" includes a region-specific loss within the Dice function. "$L_{TRS}$" denotes the extra loss items, i.e., a region-specific Tversky loss, where $\alpha$ and $\beta$ corresponds to the penalty terms for FP and FN, respectively. "T/W" is the proportion of the target regions relative to the total image area. Bold values indicate the highest DSC scores.

| Dataset | Targets | T/W (%) | Baseline | | Baseline (RS) | | Baseline + $L_{TRS}$ ($\alpha$=0.3, $\beta$=0.7) | | Baseline + $L_{TRS}$ (Adaptive $\alpha$, $\beta$) | |
|---|---|---|---|---|---|---|---|---|---|---|
| | | | DSC | Recall | DSC | Recall | DSC | Recall | DSC | Recall |
| PCD | Bronchus | 0.18 | 83.19 | 85.45 | 84.44 | 86.23 | 83.74 | 89.02 | 85.74 | 88.25 |
| | PA | 0.43 | 83.65 | 87.20 | 85.51 | 87.88 | 83.75 | 89.54 | 86.01 | 88.66 |
| | PV | 0.56 | 86.53 | 88.28 | 86.81 | 88.33 | 85.61 | 90.89 | 87.37 | 90.87 |
| | Average | 0.39 | 84.46 | 86.98 | 85.59 | 87.48 | 84.37 | 89.82 | **86.37** | 89.26 |
| LungT | Tumor | 0.13 | 68.06 | 72.83 | 71.29 | 73.37 | 70.83 | 76.36 | **72.78** | 75.48 |
| LiverT | Tumor | 0.26 | 68.77 | 79.28 | 70.52 | 80.85 | 70.49 | 87.61 | **71.37** | 88.53 |
| ABD | Duodenum | 1.65 | 66.38 | 63.73 | 70.41 | 66.52 | 69.68 | 71.23 | 72.37 | 70.76 |
| | Spleen | 2.22 | 92.21 | 91.62 | 92.82 | 92.73 | 89.72 | 93.88 | 92.30 | 94.33 |
| | Pancreas | 1.32 | 80.74 | 76.27 | 82.47 | 77.23 | 81.84 | 80.95 | 85.19 | 83.67 |
| | Aorta | 0.71 | 93.27 | 93.94 | 95.83 | 95.12 | 95.61 | 96.66 | 96.64 | 97.69 |
| | Gallbladder | 0.66 | 80.07 | 83.58 | 81.49 | 84.27 | 79.37 | 85.73 | 79.40 | 84.23 |
| | Average | 1.31 | 82.53 | 81.83 | 84.60 | 83.17 | 83.24 | 85.69 | **85.18** | 86.14 |
| BraTS | WT | 2.08 | 91.40 | 93.32 | 92.24 | 93.84 | 91.49 | 95.77 | 92.02 | 95.83 |
| | TC | 0.87 | 90.51 | 92.63 | 91.48 | 92.28 | 91.32 | 94.64 | 91.70 | 94.87 |
| | ET | 0.82 | 88.73 | 91.83 | 89.32 | 90.58 | 88.51 | 92.62 | 88.60 | 93.92 |
| | Average | 1.27 | 90.21 | 92.59 | **91.01** | 92.23 | 90.44 | 94.34 | 90.77 | 94.87 |

The experimental results demonstrate that incorporating region-specific mechanisms into the loss function improves segmentation performance. For example, in the ABD dataset, the region-specific loss increased the DSC score by 4.03 for Duodenum segmentation and by 2.56 for Aorta segmentation. When applying the Tversky loss with a heavier penalty on FN errors, we observed that recall improved but DSC slightly decreased. For instance, in the PCD dataset, recall for Pulmonary Artery (PA) segmentation increased by 1.66, while the DSC dropped by 1.76. Nevertheless, this trade-off is acceptable in clinical settings, where prioritizing recall helps avoid missing critical targets and reduces the risk of misdiagnosis. When MambaClinix adaptively balances FP and FN penalties, it achieves better optimization, improving both DSC and recall.

Additionally, we calculated the ratio of target regions to the total image area and examined its correlation with the performance of models using region-specific loss functions. Our results indicate that when this ratio is relatively small, as seen in the lung tumor and liver tumor segmentation tasks from the LungT and LiverT datasets, region-specific loss functions significantly enhance the model's learning capabilities, resulting in improved segmentation outcomes. For lung tumor segmentation, the DSC increased by 4.72, and for liver tumor segmentation, the DSC improved by 2.6. Conversely, when the target area proportion is larger, the improvement in model performance from region-specific loss functions become less pronounced. Furthermore, incorporating region-specific strategies into the Tversky loss function markedly improves recall and positively influences the DSC. Our region-specific loss is specifically designed for non-overlapping labels. Therefore, in this ablation study, for datasets with hierarchical evaluation regions (such as BraTS2021), we calculate metrics based on non-overlapping classes rather than the hierarchical regions.

## 4      Conclusion

This study proposes MambaClinix, a novel U-shaped medical image segmentation model that integrates HGCN and Mamba blocks within an adaptive stage-wise framework. Our approach further incorporates region-specific loss functions to optimize the model's decision-making capabilities. The MambaClinix architecture is capable of adaptively configuring itself to different datasets while applying region-specific strategies, positioning it as a promising tool for diverse clinical segmentation tasks. Experimental results across multiple datasets demonstrate that MambaClinix offers superior computational efficiency compared to Transformer-based methods, while significantly outperforming other state-of-the-art segmentation models.